\documentclass[showpacs,twocolumn,superscriptaddress]{revtex4}

\usepackage{doi}
\usepackage{hyperref}
\hypersetup{
  colorlinks=true,        
  linkcolor=blue,         
  citecolor=cyan,         
}

\usepackage{graphicx}
\usepackage{dcolumn}
\usepackage{bm}
\usepackage{color}

\usepackage{amsmath}
\usepackage{amssymb}

\def\a{\alpha}
\def\r{\rho}
\def\s{\sigma}
\def\t{\tau}
\def\m{\mu}
\def\n{\nu}
\def\k{\kappa}
\def\th{\theta}
\def\g{\gamma}\def\G{\Gamma}
\def\L{\Lambda}\def\l{\lambda}
\def\D{\Delta}
\def\la{\langle}
\def\ra{\rangle}
\def\o{\omega}\def\O{\Omega}
\def\d{\delta}
\def\p{\partial}

\def\half{\textstyle{\frac{1}{2}}}

\def\bdoc{\begin{document}}
\def\edoc{\end{document}}
\def\beq{\begin{equation}}
\def\eeq{\end{equation}}
\def\bea{\begin{eqnarray}}
\def\eea{\end{eqnarray}}
\def\ben{\begin{enumerate}}
\def\een{\end{enumerate}}
\def\la{\langle}
\def\ra{\rangle}
\def\a{\alpha}
\def\b{\beta}
\def\g{\gamma}
\def\G{\Gamma}
\def\d{\delta}
\def\D{\Delta}
\def\e{\epsilon}

\def\th{\theta}
\def\k{\kappa}
\def\l{\lambda}
\def\m{\mu}
\def\n{\nu}
\def\o{\omega}
\def\p{\pi}
\def\r{\rho}
\def\s{\sigma}
\def\t{\tau}
\def\L{{\cal L}}
\def\S{\Sigma }
\def\gsim{\; \raisebox{-.8ex}{$\stackrel{\textstyle >}{\sim}$}\;}
\def\lsim{\; \raisebox{-.8ex}{$\stackrel{\textstyle <}{\sim}$}\;}
\def\gtrsim{\gsim}
\def\lessim{\lsim}
\def\loc{{\rm local}}
\def\vm{v_{\rm max}}
\def\bh{\bar{h}}
\def\del{\partial}
\def\nab{\nabla}
\def\half{{\textstyle{\frac{1}{2}}}}
\def\fourth{{\textstyle{\frac{1}{4}}}}

\def\bD{{\bf D}}
\def\bE{{\bf E}}
\def\bF{{\bf F}}
\def\bB{{\bf B}}
\def\bP{{\bf P}}
\def\bV{{\bf v}}
\def\bv{{\bf v}}
\def\bx{{\bf x}}
\def\by{{\bf y}}
\def\bz{{\bf z}}
\def\ba{{\bf a}}
\def\bd{{\bf d}}
\def\bs{{\bf s}}
\def\bn{{\bf n}}
\def\bp{{\bf p}}

\def\O{\Omega}

\def\br{{\bf r}}
\def\bnab{{\bf \nab}}

\def\tE{\tilde{E}}
\def\tL{\tilde{L}}

\begin{document}

\title{Motion of Particles and Gravitational Lensing Around (2+1)-dimensional BTZ black holes in Gauss-Bonnet Gravity}

\author{Bakhtiyor Narzilloev}
\email[]{nbakhtiyor18@fudan.edu.cn}
\affiliation{Center for Field Theory and Particle Physics and Department of Physics, Fudan University, 200438 Shanghai, China }
\affiliation{Akfa University, Kichik Halqa Yuli Street 17, Tashkent 100095, Uzbekistan}
\affiliation{Ulugh Beg Astronomical Institute, Astronomy St 33, Tashkent 100052, Uzbekistan}

\author{Sanjar Shaymatov}
\email[]{sanjar@astrin.uz}

\affiliation{Ulugh Beg Astronomical Institute, Astronomy St 33, Tashkent 100052, Uzbekistan}
\affiliation{Akfa University, Kichik Halqa Yuli Street 17, Tashkent 100095, Uzbekistan}
\affiliation{Institute for Theoretical Physics and Cosmology, Zheijiang University of Technology, Hangzhou 310023, China}
\affiliation{National University of Uzbekistan, Tashkent 100174, Uzbekistan}
\affiliation{Tashkent Institute of Irrigation and Agricultural Mechanization Engineers, Kori Niyoziy 39, Tashkent 100000, Uzbekistan}

\author{Ibrar Hussain}
\email[]{ibrar.hussain@seecs.nust.edu.pk}

\affiliation{School of Electrical Engineering and Computer Science, National University of Sciences and Technology, H-12, Islamabad, Pakistan}

\author{Ahmadjon~Abdujabbarov}
\email[]{ahmadjon@astrin.uz}
\affiliation{Ulugh Beg Astronomical Institute, Astronomy St 33, Tashkent 100052, Uzbekistan}
\affiliation{Institute of Nuclear Physics, Ulugbek 1, Tashkent 100214, Uzbekistan}
\affiliation{National University of Uzbekistan, Tashkent 100174, Uzbekistan}
\affiliation{Tashkent Institute of Irrigation and Agricultural Mechanization Engineers, Kori Niyoziy 39, Tashkent 100000, Uzbekistan}
\affiliation{Shanghai Astronomical Observatory, 80 Nandan Road, Shanghai 200030, P. R. China}

\author{Bobomurat Ahmedov}
\email[]{ahmedov@astrin.uz}

\affiliation{Ulugh Beg Astronomical Institute, Astronomy St 33, Tashkent 100052, Uzbekistan}
\affiliation{National University of Uzbekistan, Tashkent 100174, Uzbekistan}
\affiliation{Tashkent Institute of Irrigation and Agricultural Mechanization Engineers, Kori Niyoziy 39, Tashkent 100000, Uzbekistan}

\author{Cosimo Bambi}
\email[]{bambi@fudan.edu.cn}
\affiliation{Center for Field Theory and Particle Physics and Department of Physics, Fudan University, 200438 Shanghai, China }

\date{\today}
\begin{abstract}
We study motion of test particles and photons in the vicinity of (2+1) dimensional Gauss-Bonnet (GB) BTZ black hole. We find that the presence of the coupling constant serves as an attractive gravitational charge, shifting the innermost stable circular orbits outward with respect to the one for this theory in 4 dimensions. Further we consider the  gravitational lensing, to test the GB gravity in (2+1) dimensions and show that the presence of GB parameter causes the bending angle to grow up first with the increase of the inverse of closest approach distance, $u_0$, then have its maximum value for specific $u_0^*$, and then reduce until zero. We also show that increase in the value of the GB parameter makes the bending angle smaller and the increase in the absolute value of the negative cosmological constant produces opposite effect on this angle.
\end{abstract}
\maketitle


\section{Introduction}
\label{introduction}

The existing large amount of data from observations of astrophysical processes around compact objects, such as gravitational waves~\cite{LIGO16}, black hole shadows~\cite{EHT19a,EHT19b},  etc., together with gravity theories can provide useful tool to understand the nature of the gravitational interaction~\cite{Yagi:2016jml,Cardoso:2016ryw,Bambi:2015kza}. On the other hand testing and constraining the parameters of the gravity models is a step forward to discover the unified field theory. 

Avoiding the Lovelock's theorem indicating that in less than 5 dimensions the cosmological constant can appear only within the general relativity~\cite{Lovelock71}  recently there has been proposed a new approach to obtain the solution in the Einstein-Gauss-Bonnet (EGB) gravity in 4 and 3 dimensional $(D=4, D=3)$ spaceimes~\cite{Glavan20}. The approach based on using the rescaling of the Gauss-Bonnet term in such a way that the limit $D\rightarrow 4$ ($D\rightarrow 3$ ) does not diverge. 

A lot of work has been done in the literature on EGB gravity in $D=4$. Particularly, authors of Ref.~\cite{Malafarina20} have studied the gravitational collapse in 4-D EGB gravity and have  showed the similarity of spherical dust collapse to one in Einstein's gravity. The effect of the GB coupling constant on the superradiance in black hole spacetimes have been studied in~\cite{Zhang20a}. The stability of linearized equations of motion has been analyzed in Ref.~\cite{Aguilar19}, exploring the perturbations of the black hole event horizon in the GB gravity. The Ref.~\cite{Zhang20} is devoted to study the classical spinning test particle motion around non-rotating black hole in a 4-D EGB gravity. Authors of Ref.~\cite{Guo20} have analyzed the motion of test particle along the geodesic around 4-D EGB black hole. Charged particle and epicyclic motions around 4-D EGB black hole immersed in an external magnetic field has also been considered in Ref.~\cite{Shaymatov20egb}.  

The rotating analogue of static black hole solution in 4-D EGB gravity has been obtained in~\cite{Kumar20}. Analyzing the scalar and electromagnetic perturbation around 4-D EGB rotating black hole, the authors of Ref.~\cite{Mishra20} have checked the strong cosmic censorship conjecture. Other properties of the rotating spacetime in 4-D EGB black hole have been explored in  Refs.~\cite{Churilova20,Aragon20,Islam20,Dadhich20egb,Liu20egb}. The energetics and shadow of higher dimensional i.e 6-D EGB black hole have been studied in~\cite{Abdujabbarov15a}. Also the authors of Refs.~\cite{Shaymatov20-pl,Dadhich21} have investigated overspinning of 6-D EGB black hole and circular orbits around higher dimensional Einstein and Gauss-Bonnet rotating black holes.

The thermodynamics, phase transition and Joule-Thomson expansion of 4-D Gauss-Bonnet AdS black hole~\cite{Hegde20}, particle acceleration~\cite{Kumara20}, thermodynamic geometry of the 4-D EGB AdS black hole~\cite{Mansoori20}, the emergent universe scenario in the 4-D EGB gravity~\cite{Li20a}, extended thermodynamics and microstructures of 4-D charged EGB black hole~\cite{Wei20}, the shadow of rotating 4-D EGB black hole~\cite{Wei20a} and  gravitational lensing of 4-D EGB black hole~\cite{Heydari20,Atamurotov21a} have been widely studied in the literature. 

At the same time there has been a discussion of the validity of the model.  Particularly, Ref.~\cite{Gurses20} has discussed the problem of existance of EGB theory in 4-D spacetime. The main conclusion of Ref.~\cite{Hennigar20} was that 4-D EGB is not well defined. Other authors~\cite{Arrechea20,Tian20} have also questioned the validity of the field equations in the case of 4-D EGB gravity. At the same time in the limiting case when $D\rightarrow 4$ the higher-dimensional scattering amplitudes of the GB theory differs from the general relativity and may be caused by additional scalar-tensor field~\cite{Bonifacio20}.  
\textcolor{black}{Here we plan to study the spacetime properties around the 3-D Gauss-Bonnet BTZ black hole using the analysis of test particles and photon motion. This study may be useful for developing new tests of the 3-D Gauss-Bonnet theory and get corresponding constraints on parameters of the theory.} Another interesting aspect of studying black holes in dimension $D < 4$ is that they may refuse what is true for black holes in dimension $D=4$~\cite{Duztas-Jamil20}. We use the solution obtained in Ref.~\cite{Hennigar20PLB} describing the 3-Ds BTZ black hole in the EGB theory gravity. 

The test particle motion as well as photon trajectories are useful tool to explore the spacetime properties and its structure in various gravity theories~\cite{Banados09,Babar16,Shaymatov13,
Zakria15,Jawad16,Jamil15,Hussain17, Brevik19, Shaymatov20a, DeLaurentis2018PhRvD, Shaymatov21a,Atamurotov13a, Hakimov17, Narzilloev19, Narzilloev:2020b, Narzilloev20c, Narzilloev20a, Narzilloev20b,Shaymatov21b, Shaymatov21c, Narzilloev21, Narzilloev21b}. Particularly the charged and magnetized particles motion around black hole in the presence of external magnetic field have been widely studied in Refs.~\cite{deFelice, defelice2004,Wald74,Shaymatov14,Shaymatov15,Shaymatov18,Aliev02,Aliev2004,Aliev05,Aliev06,Frolov10,Abdujabbarov10,Abdujabbarov11a,Frolov12,Karas12a,shaymatov19b,Stuchlik14,Stuchlik16,Shaymatov20b,Stuchlik20,Shaymatov21c}. 

Photon motion and its deflection in the gravitational field is one of the main features of the metric theories of gravity. One may read the review of the effects of gravitational lensing in Refs.~\cite{Synge60,Schneider92,Perlick00,Perlick04}. 
A number of works have been devoted to explore the gravitational lensing in the weak and strong field regimes~\cite{Rogers15,Rogers17,Er18,Rogers17a,Broderick03,Bicak75, Kichenassamy85,Perlick17,Perlick15,Abdujabbarov17,Eiroa12, Kogan10,Tsupko10,Tsupko12,Morozova13,Tsupko14,Kogan17,Benavides16,Kraniotis14}. One of the consequences of the exploration of photon motion leads to shadow of the black holes. The discovery of the image of the black hole shadow by EHT team~\cite{EHT19a,EHT19b} has been conducted with the theoretical studies of this effect by various authors~\cite{Takahashi05, Hioki09, Bambi09, Bambi10, Amarilla10, Amarilla12, Amarilla13, Abdujabbarov13c, Bambi13c, Wei13, Li14a, Bambi15, Ghasemi-Nodehi15, Cunha15, Abdujabbarov15, Atamurotov15a, Ohgami15, Grenzebach15, Mureika17, Abdujabbarov17b, Abdujabbarov16b, Mizuno18, Shaikh18b, Schee15, Schee09a, Schee09, Stuchlik10, Bambi:2019tjh, Hensh19}. In the present study we are keen to explore the gravitational lensing in the 3-D EGB gravity. In particular we examine the effects of the GB parameter and of the cosmological constant on the gravitational lensing in the the 3-D EGB gravity. 

The paper is organized as follows: The Sec.~\ref{Sec:metic} is devoted to the review of the 3-D BTZ black hole solution in EGB theory. We study the test particle motion around the 3-D BTZ black hole in EGB theory in Sec.~\ref{Sec:motion}. We explore the photon motion and gravitational lensing in Sect.~\ref{sec:phot}. We conclude our discussion in Sec.~\ref{Sec:conclusion}. 
Throughout this work we use a system of units in which $G=c=1$. Greek indices are taken to run from 0 to 2, while Latin ones from 1 to 2.
\section{\label{Sec:metic} 3D Gauss-Bonnet BTZ black hole metric }

In $D$ dimensions, the action for the GB theory with scalar field $\phi$ is given by
\begin{eqnarray}
S&=&\frac{1}{16\pi G}\int\sqrt{-g} d^Dx\left[R-2\Lambda+\alpha \left(\phi L_{GB}\right.\right.\nonumber\\&+&\left.\left. 4G^{\mu\nu}\partial_{\mu}\phi\partial_{\nu}\phi-4\left(\partial\phi\right)^2\square\phi+2\big(\left(\triangledown\phi\right)^2\big)^2 \right)\right]\, ,
\end{eqnarray}
where $\alpha$ refers to the the the GB coupling constant and $L_{GB}$ the GB term  is given by  
\begin{eqnarray}
L_{GB}=R_{\mu\nu\lambda\delta}R^{\mu\nu\lambda
	\delta} - 4R_{\mu\nu}R^{\mu\nu}+R^2\, ,
\end{eqnarray}
with $R$ being the scalar curvature. Recently \cite{Glavan20} proposed new approach that it is possible to obtain GB contribution in $D=4$ by rescaling the coupling constant. It has also been shown \cite{Hennigar20PLB}, that it is possible to obtain $D=3$ case of the theory with $\phi=\ln(\frac{r}{\ell})$, where $\ell$ is the constant of integration. It is worth noting that~\cite{Hennigar20PLB,Hennigar20gb} for $D=3$ the GB term $L_{GB}$ vanishes.  

The form of the GB theory in $D=3$ is given by \begin{eqnarray}\label{solution}
ds^2=-F(r)dt^2+\frac{dr^2}{F(r)}+r^2d\varphi^2,
\end{eqnarray}
with
\begin{eqnarray}
F(r)=-\frac{r^2}{2\alpha}\left(1\pm\sqrt{1+\frac{4\alpha }{r^2}f_{E}}\right),
\end{eqnarray}
where 
\begin{eqnarray}
f_{E}=\frac{r^2}{l^2}-m\, ,
\end{eqnarray}
where $m$ and $l$ respectively refer to the integration constants with $\Lambda=-1/l^2$. Note that  this $3-D$ GB theory has two separate black hole solutions, i.e. $\pm \sqrt{1+\frac{4\alpha }{r^2}f_{E}}$. 
It is worth noticing that in the spacetime of 3D BTZ black hole in GB gravity one can calculate the effective AdS radius from the condition $F(r)=0$ that reads as
\begin{eqnarray}\label{rads}
r_{AdS}=\sqrt{-\frac{m}{\Lambda}}\, ,
\end{eqnarray}
which is independent of the GB coupling parameter, $\alpha$, while in higher-curvature theories this radius is a function of the coupling parameter~\cite{Shaymatov20egb,Dadhich20egb,Liu20egb,Kumar20,Guo20,Wei20a}. From the above equation, it is clearly seen that the integration constant $m$ is positive quantity as $\Lambda$ takes negative values only. Note that this AdS radius coincides with one for 3D BTZ black hole in Einstein gravity in the limit $J\to 0$ ~\cite{Cruz94BTZ}. It can be also seen that the spacetime metric allows the cosmological constant to take only negative values to have real AdS radius corresponding to the cosmological horizon which tends to infinity in the case in which  $\Lambda\rightarrow0$.

Here we restrict ourselves to the 'minus' branch of solution as it leads to a well-defined BTZ solution only in the limiting case of small $\alpha$,  
i.e., when the metric function $F(r)$ is expanded in the form~\cite{Hennigar20PLB}
\begin{equation}
\lim_{\alpha\rightarrow 0} F(r) = \frac{r^2}{l^2}-m-\frac{\alpha}{r^2}\left(\frac{r^2}{l^2}-m\right)^2+O(\alpha^2)\, ,
\end{equation}
which coincides with the one for standard BTZ solution for small $\alpha$ in the Einstein gravity. However,
in the limiting case of large $r$ the metric function $F(r)$ yields \begin{equation}
F(r) = \frac{r^2}{2\alpha}\left(\sqrt{1+\frac{4\alpha}{l^2}}-1\right)-\frac{m}{\sqrt{1+\frac{4\alpha}{l^2}}}+O(1/r^2)\, .
\end{equation} 
From the above expression, the coupling parameter can take negative values, i.e. $\alpha>-l^2/4$ at the large distances (see for example~\cite{Hennigar20PLB} ). However, this would not be the case in close vicinity of the 3D BTZ black hole. Therefore, we further consider positive values of $\alpha$ to explore the properties of 3D BTZ black hole in GB gravity.

\section{\label{Sec:motion} Test particles motion }

Here we consider a test particle motion in the gravitational field of the (2+1)-dimensional BTZ black hole in the GB gravity. 
\begin{figure*}
	\centering
	\includegraphics[width=0.46\textwidth]{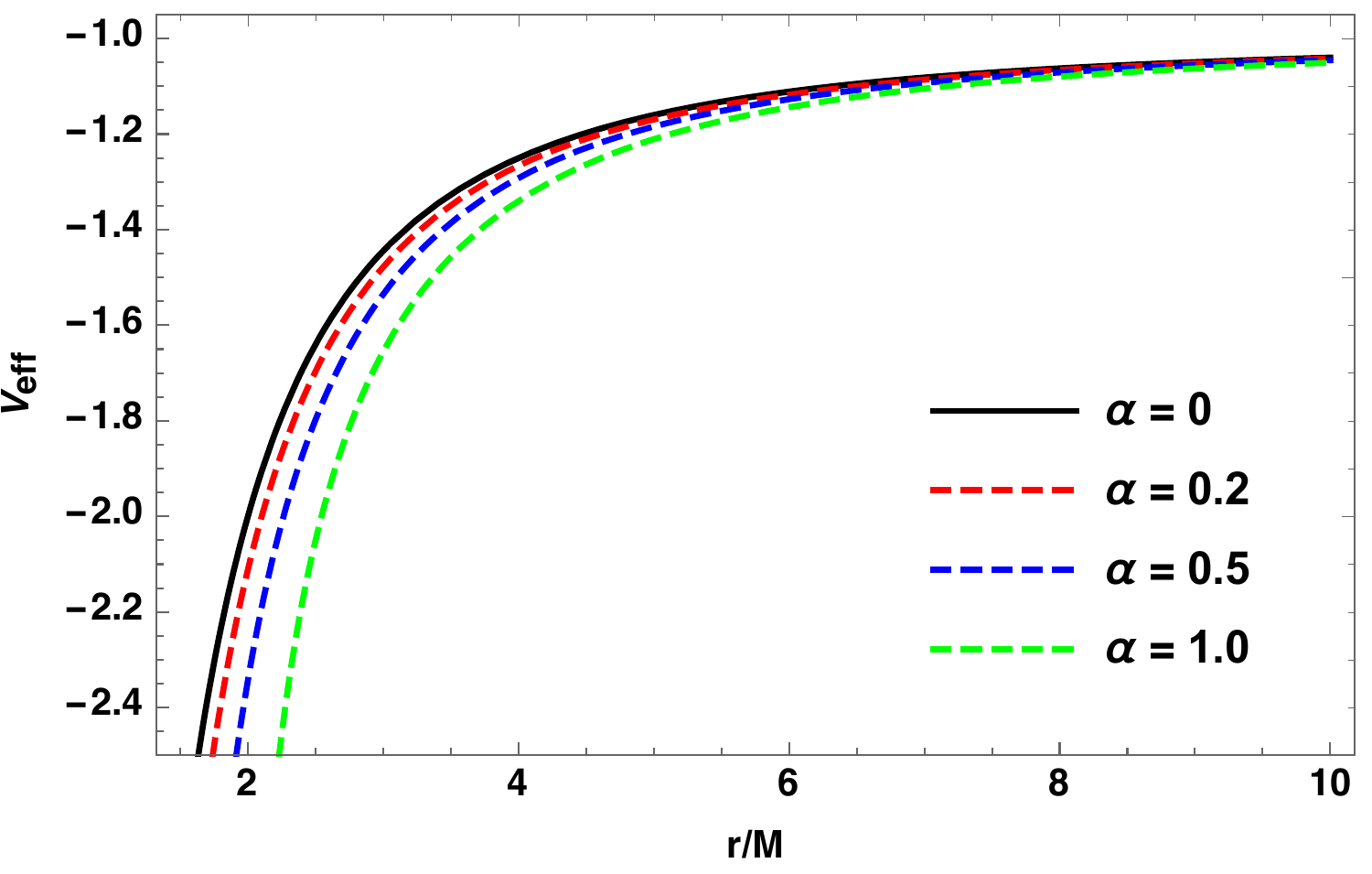}
	\includegraphics[width=0.45\textwidth]{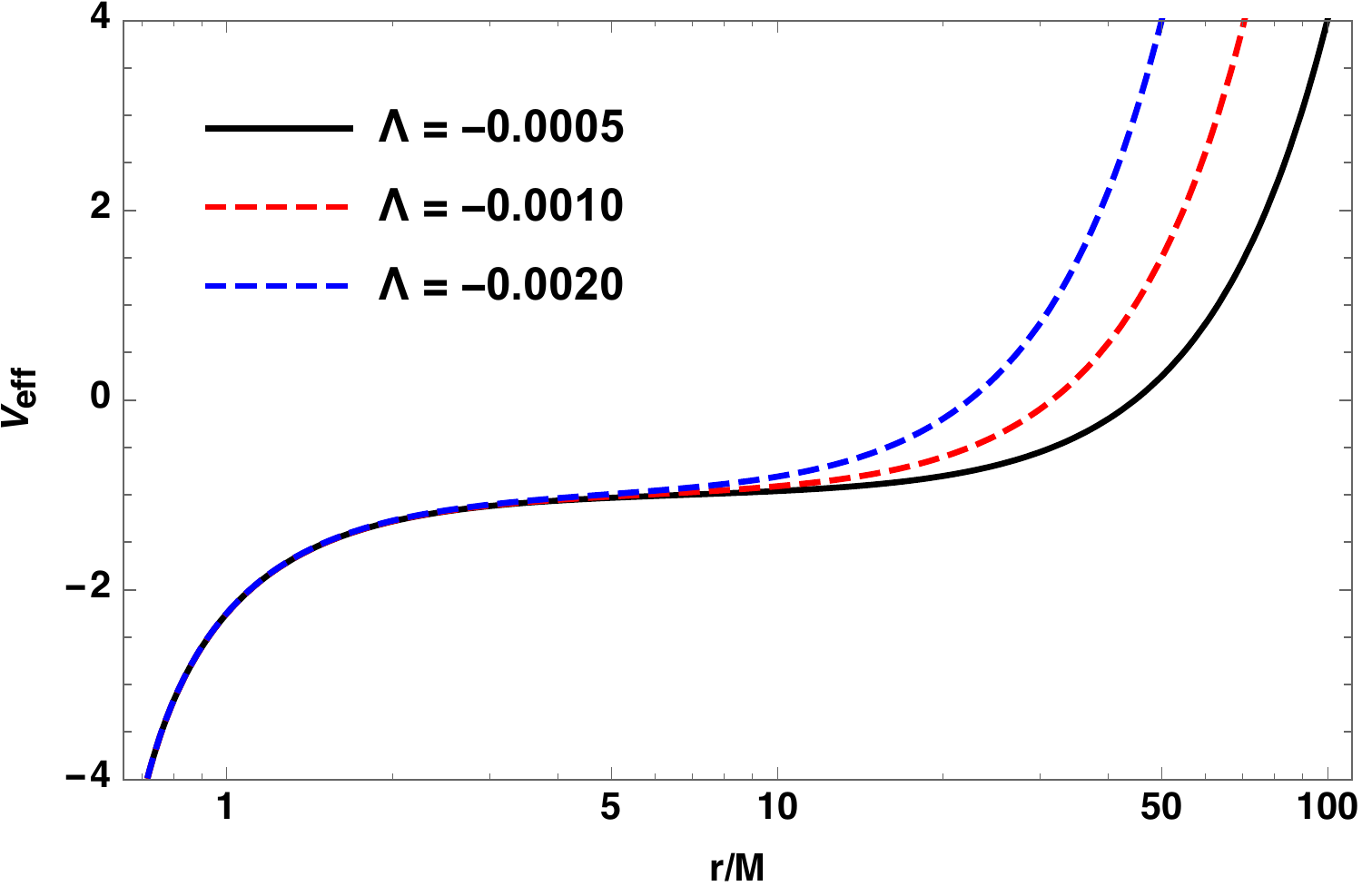}
	
	\caption{\label{fig1} Radial dependence of the effective potential for massive particles moving around GB (2+1)-dimensional BTZ black hole. $V_{eff}$ is plotted for different values of GB parameter $\alpha$ for given $\Lambda=0$ in the left panel while for different values of $\Lambda$ for given $\alpha=0.1$ in the right panel. } \end{figure*}
\begin{figure*}
	\centering
	\includegraphics[width=0.45\textwidth]{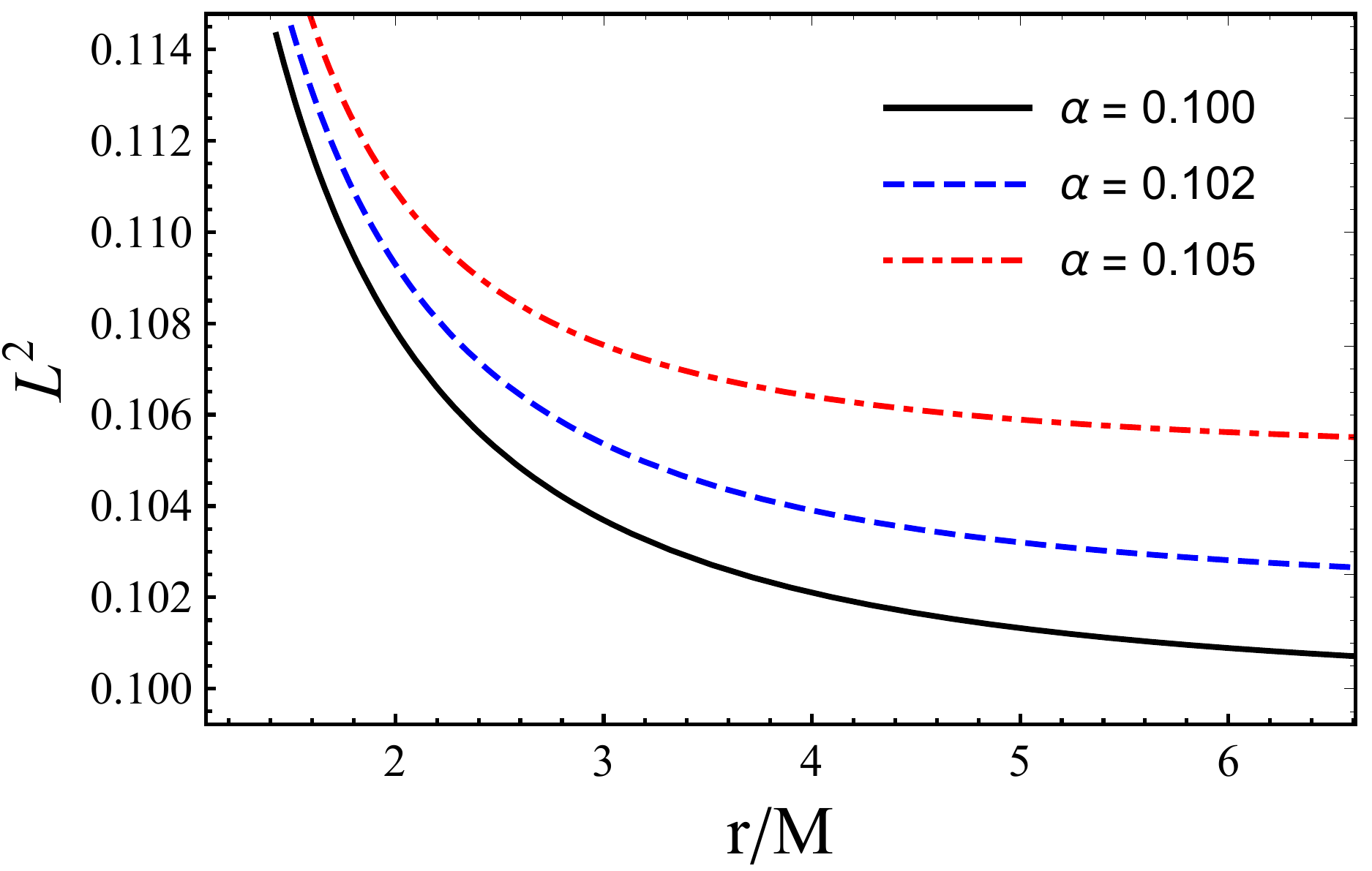}
	\includegraphics[width=0.45\textwidth]{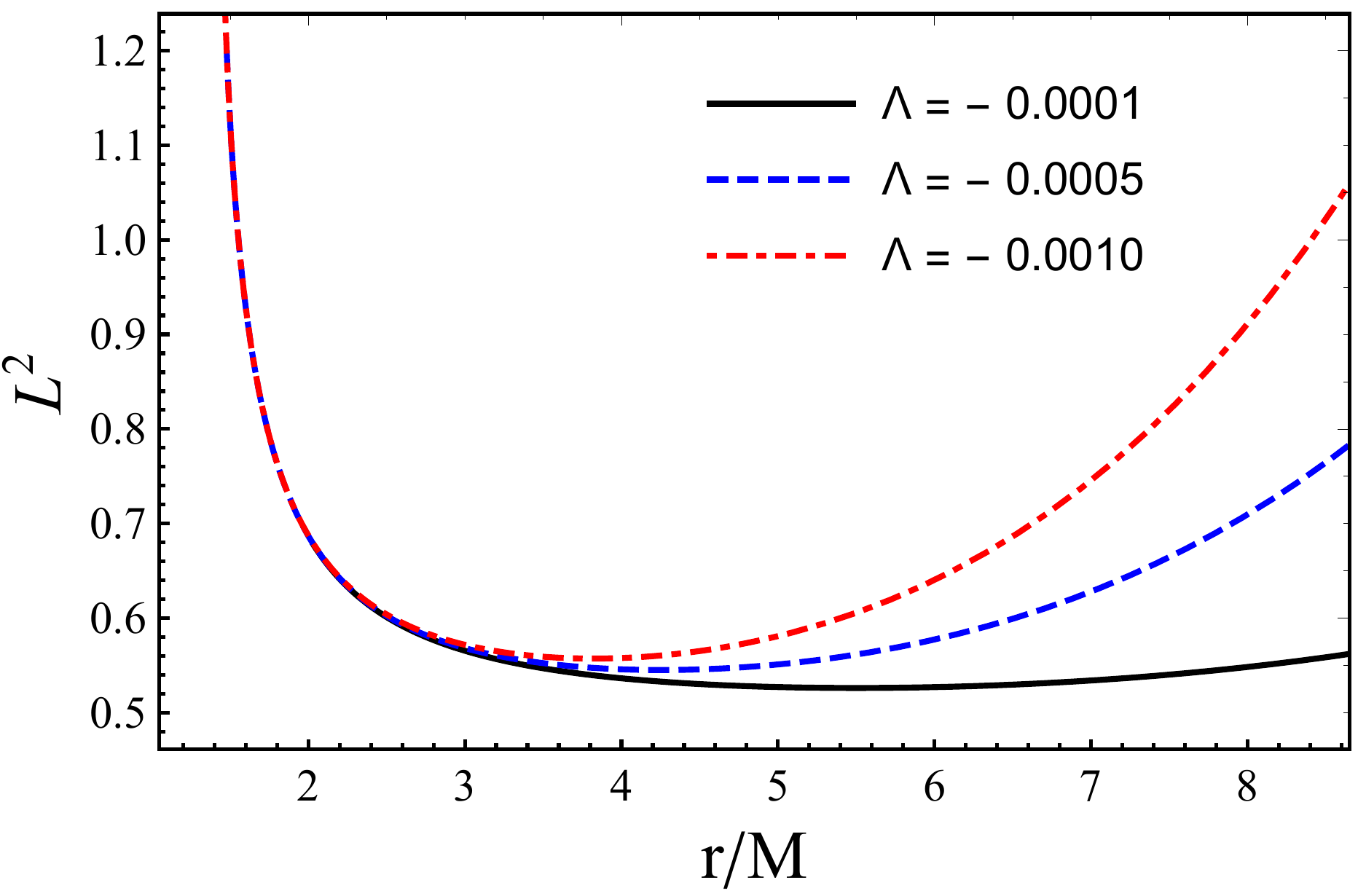}
	
	\caption{\label{fig2} The radial dependence of the specific angular momentum for test particles around GB (2+1) -dimensional BTZ black hole. Left panel: for the different values of parameter $\alpha$ in the case of vanishing cosmological constant, i.e. $\Lambda=0$.  Right panel: for  different values of cosmological constant in the case of fixed $\alpha=0.5$.  } \end{figure*}
To study the motion of test particles in the vicinity of the BTZ black hole we explore the Hamilton-Jacobi equation~\cite{Misner73} 
\begin{eqnarray}
H  \equiv \frac{1}{2}g^{\mu\nu}\left(\frac{\partial S}{\partial x^{\mu}} \right)\left(\frac{\partial S}{\partial x^{\nu}} \right)\, ,
\label{Eq:H}
\end{eqnarray}
with the action $S$ and the coordinate three-vector $x^\mu$. The Hamiltonian is a constant that can be set to $H=k/2$ with $k=-m^{\prime 2}$ (where $m^{\prime}$ is the mass of the test particle).  

Then from the symmetry of the system one can write the action $S$ for the motion of test particles around the black hole in separable form  as 
\begin{eqnarray}\label{Eq:action}
S= \frac{1}{2}k\lambda-Et+L\varphi+S_{r}(r)\ .
\end{eqnarray}
Here $E$ and $L$ refer to the energy and angular momentum of the particle,  respectively, and the parameter $\lambda$ is an affine parameter. From Eq. (\ref{Eq:action}), we rewrite Hamilton-Jacobi equation in the following form 
\begin{eqnarray}\label{Eq:separable}
&k=& -F(r)^{-1}E^2
+ F(r)
\left(\frac{\partial S_{r}}{\partial r}\right)^2
+ \frac{L^2}{r^2} \, , 
\end{eqnarray}
In the (2+1)-dimensional system there appear three independent constants of motion, i.e. $E$, $L$ and $k$ which have been specified in~\cite{Misner73}. In this case the corresponding constant related to the latitudinal motion of particles is not available as that of the properties of the BTZ spacetime. 
From Eq.~(\ref{Eq:separable}) we obtain the radial equation of motion for particles as 
\begin{eqnarray}
\frac{1}{2}\dot{r}^{2} + V_{eff}(r;\mathcal{L},\alpha,\Lambda)=\mathcal{E}^2\, ,
\end{eqnarray}
where the dot denotes derivative with respect to the proper time $\tau$ and the radial function $V_{eff}(r;\mathcal{L},\alpha,\beta)$ refers to the effective potential of the system which is given by 
\begin{eqnarray} \label{Eq:Veff}
V_{eff}(r;\mathcal{L},\alpha,\Lambda) &=&  \frac{r^2}{2\alpha}\left(\sqrt{1-\frac{4\alpha \left(\Lambda r^2 +m\right)}{r^2}}-1\right)\nonumber\\&\times&\left(
1+\frac{\mathcal{L}^2}{r^2}\right)\, , 
\end{eqnarray}
with the conserved constants per unit mass $m^{\prime}$ given by $\mathcal{E}=E/m^{\prime}$ and $\mathcal{L}=L/m^{\prime}$ and $k/m^{\prime 2}=-1$.

In Fig.~\ref{fig1} we demonstrate the radial dependence of the effective
potential (\ref{Eq:Veff}) for different values of GB parameter $\alpha$ and cosmological constant $\Lambda$. From Fig.~\ref{fig1}, with increasing $\alpha$ the curves start coming down. However, we can see that the negative cosmological constant, i.e. $\Lambda<0$ has the opposite
effect with respect to the GB parameter $\alpha$, thereby suggesting that the effect of the cosmological constant can prevent test particles from escaping or falling into the black hole. Since the situation gets altered for cosmological constant test particles under the effect of $\alpha$ and $\Lambda$ can have stable circular orbits around (2+1)-dimensional BTZ black hole in the GB gravity. 

Following effective potential (\ref{Eq:Veff}), we turn to the study of stable circular orbits of test particles around BTZ black hole in the GB gravity.  For test particles to be on stable circular orbits we shall focus on  the required conditions for which  
\begin{eqnarray}\label{Eq:cir1}
V_{ eff}(r;\mathcal{L},\alpha,\Lambda)=\mathcal{E}^2\ , \\ \nonumber\\ \frac{\partial V_{eff}(r;\mathcal{L},\alpha,\Lambda)}{\partial r}=0\, .
\label{Eq:cir2}
\end{eqnarray}
From the above equations the corresponding energy and angular momentum for test particles on stable circular orbits can be obtained. Further we show the dependence on the GB parameter and cosmological constant of the angular momentum required for the test particles to be on stable circular orbit around the black hole, see Fig.~\ref{fig2}. It is clearly shown in Fig.~\ref{fig2} that stable circular orbit of particles shifts toward the large radii as a result of an increase in the value of $\alpha$. On the other hand, large values of $\alpha$ give rise to the increase in the value of $\mathcal{L}$, showing that particle to stay on stable circular orbit needs more angular momentum for biger $\alpha$. The situation however gets altered as the cosmological parameter has the opposite effect, thereby reducing the value of the angular momentum for the particle to be on stable circular orbit around (2+1)-dimensional BTZ black hole in the GB gravity.
\begin{table}
\caption{\label{tab1} The values of the ISCO radius $r_{isco}$ are tabulated in the case of massive particles moving around $3-D$ GB black hole for different values of GB parameter $\alpha$ and cosmological constant $\Lambda$. }
\begin{tabular}{c|cccc}
$$  & $$ &  $$
& $\Lambda$ & $$  \\ \hline
{$\rm \alpha $}   & $-0.0001$ &  $-0.0005$
& $-0.0010$ & $-0.0050$  \\
\hline\\
0.01    &1.01000  &0.77779  &0.695958 &0.540149  \\\\
0.05    &1.73920 &1.34584 &1.20781 &0.94661  \\\\
0.1    &2.20810 &1.70810 &1.53567 &1.21051  \\\\
0.5    &3.81943 &2.99347 &2.70679 &2.17044  \\
\end{tabular}
\end{table}
\begin{figure}
	\centering
	\includegraphics[width=0.45\textwidth]{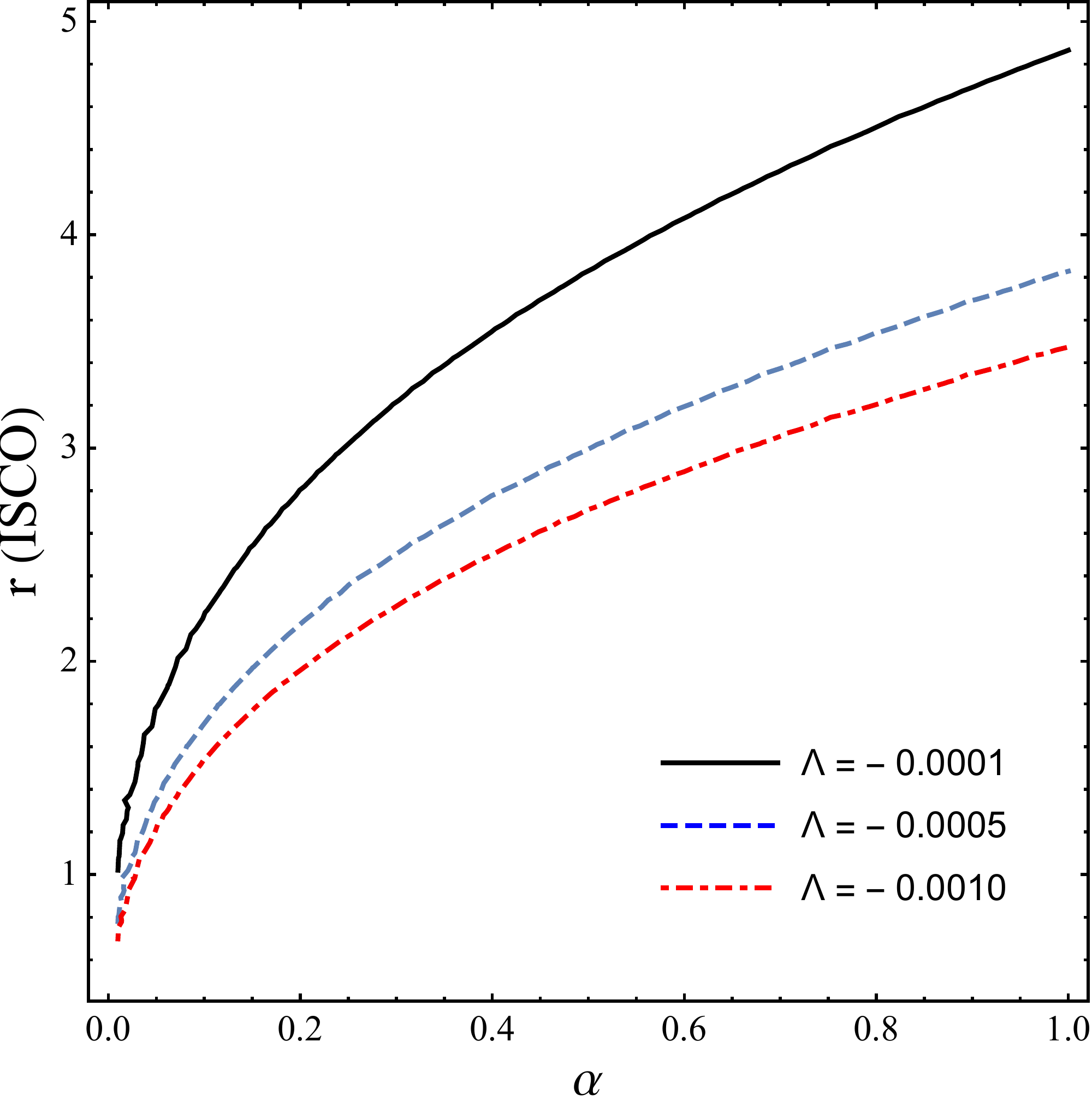}
	
	\caption{\label{fig3} The dependence of the ISCO radius on the GB parameter $\alpha$ for different values of cosmological constant $\Lambda$.  } 
	\end{figure}
\begin{figure*}
	\centering
	\includegraphics[width=0.45\textwidth]{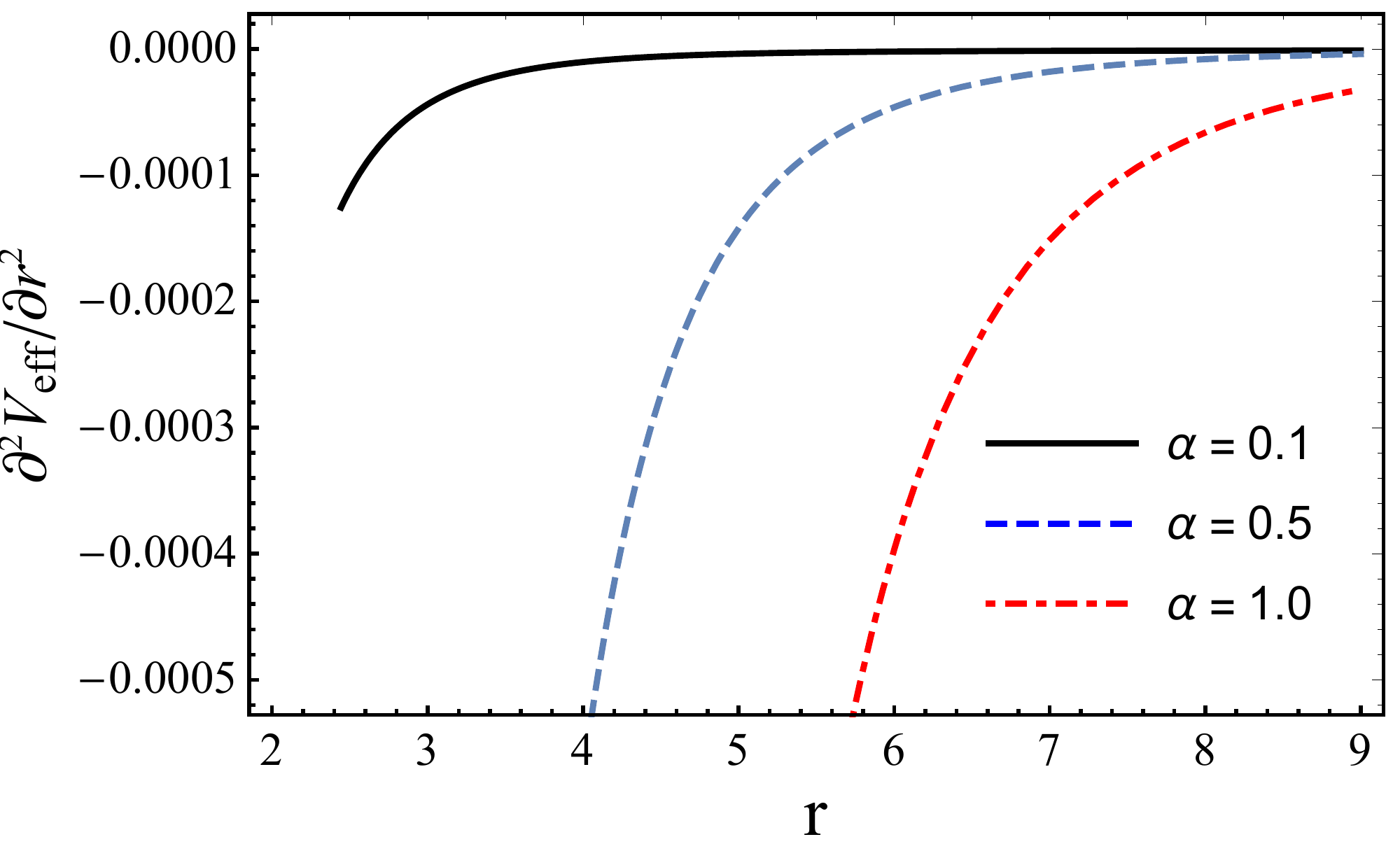}
	\includegraphics[width=0.45\textwidth]{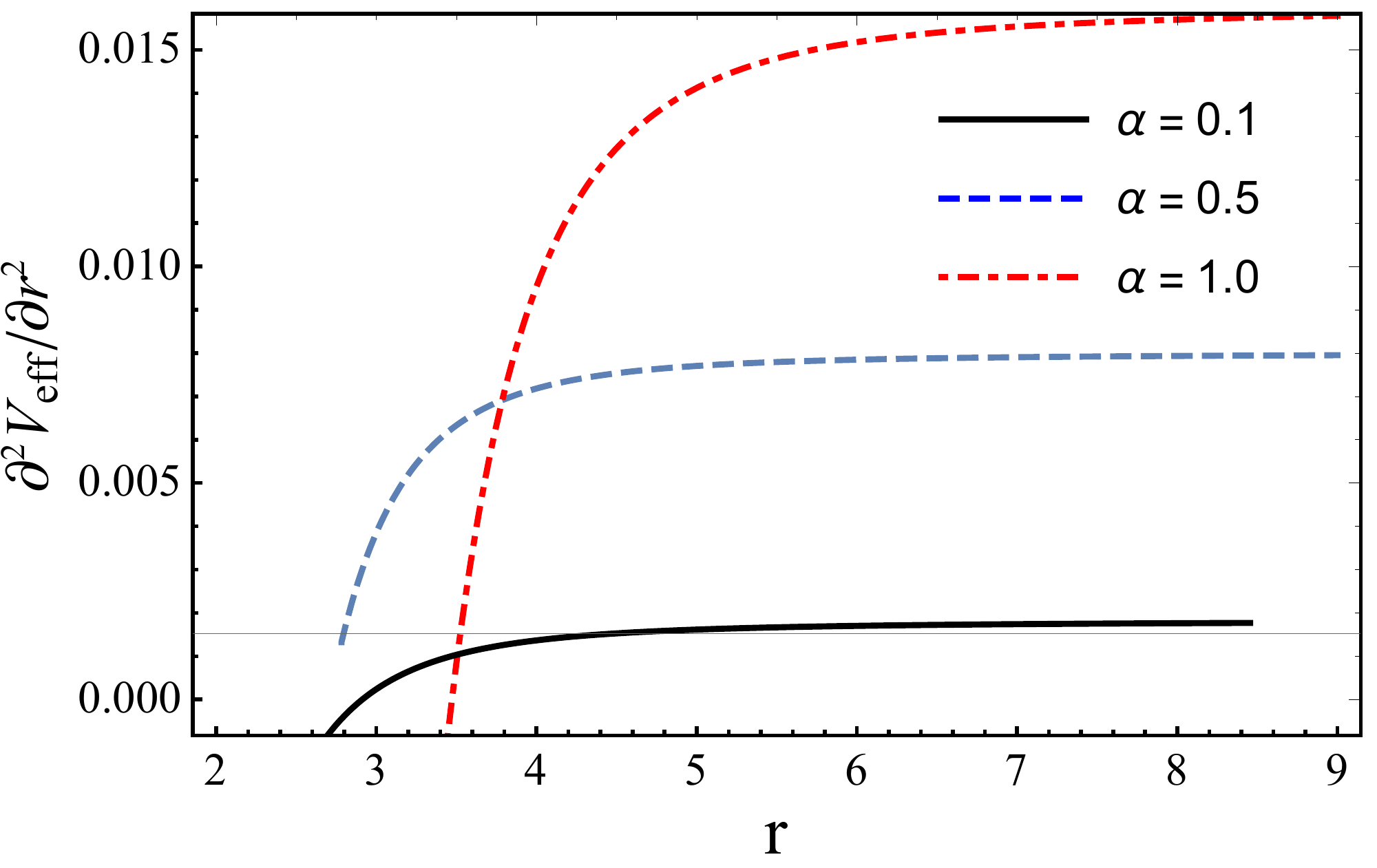}
	
	\caption{\label{fig4} Radial profile of $\partial^2 V_{eff}/\partial r^2$ for test particles making circular motion around (2+1) -dimensional BTZ black hole in the GB gravity is plotted for different values of GB parameter $\alpha$. Left/right panels refer to $\Lambda=0/- 0.001$. \textcolor{black}{Whereas, in the case of (2+1) -dimensional BTZ black hole in Einstein gravity the situation gets totally altered from the one in GB gravity where $V_{eff}^{\prime\prime}$ does not depend upon $r$, ( $V_{eff}^{\prime\prime}=-8\Lambda$) and hence there occur no stable circular orbits for massive particles~(see for example \cite{Cruz94BTZ}). } } \end{figure*}
	
Further we explore the innermost stable circular orbits (ISCO), which is given by the auxiliary condition on the effective potential
\begin{eqnarray}\label{eq:isco}
 \frac{\partial^2 V_{eff}(r;\mathcal{L},\alpha,\Lambda)}{\partial r^2}=0\, .
\end{eqnarray}
In the case of massive particles the radius of the ISCO $r_i$ is obtained from the minimum value of the angular momentum $\mathcal{L}$ determined by ${ V_{eff}^{\prime}(r;\mathcal{L},\alpha,\Lambda)}=0$.
For the existence of innermost stable circular orbits the above condition must always be satisfied. It is difficult to solve and analyse Eq.~(\ref{eq:isco}) analytically, and hence  we explore the ISCO radius numerically (see Table~\ref{tab1}).
As can be seen from the Table~\ref{tab1}, the ISCO radius increases with increasing the GB parameter $\alpha$ while it decreases with increasing the cosmological parameter $\Lambda$. This behaviour of the ISCO radius is also shown very clearly in Fig.~\ref{fig3}. \textcolor{black}{ From Eq.~\eqref{rads} it is noticeable that the AdS radius corresponding to the cosmological horizon is always greater than the radius of stable circular orbits of the test particles for the chosen values of the cosmological constant.}
It turns out that the resultant gravitational force becomes stronger as we increase GB parameter, thereby indicating that it acts as an attractive charge, whereas an increase in the value of the cosmological constant weakens the gravity. Thus, for massive particle to be under the combined effect of GB parameter and cosmological constant it makes sense so as to move at stable circular orbits around (2+1)-dimensional BTZ black hole in the GB gravity. This leads to an interesting question what happens to massive particle in case the effect of cosmological parameter is switched off --could it be fallen into the black hole under the effect of GB parameter?  To settle this question we shall analyse Eq.~(\ref{eq:isco}), whether it is always negative or not. With this aim, we plot the radial profile of $V_{eff}^{\prime\prime}(r)$ for various values of GB parameter, see Fig.~\ref{fig4}. We note that in the case of $\Lambda=0$, $V_{eff}^{\prime\prime}(r)< 0$ always at all values of $r$, thereby indicating that there occurs no stable circular orbits around black hole having GB parameter only. The situations however gets overturned once the effect arising from the cosmological constant is included, showing that $V_{eff}^{\prime\prime}(r)> 0$ always at all $r$, see (Fig.~\ref{fig4} right panel). Or, in other words, the particles can be on stable circular orbits around (2+1)-dimensional BTZ black hole in the GB gravity. \textcolor{black}{It is interesting that this behaviour for stable circular orbits exhibits striking difference from its Einstein counterpart for which there exists no occurrence of stable orbits~\cite{Cruz94BTZ}.  However, in the GB gravity case the cosmological constant plays a crucial role in getting information of stable circular orbits of test particles around BTZ black hole and its properties as well.}   
	
\section{Photon motion and gravitational lensing \label{sec:phot}}

In this part we consider the motion of photon in the GB (2+1)-dimensional BTZ black hole spacetime. From the usual relation for the 3-momentum (note that we are dealing with 3D spacetime) $p_\mu p^\mu=k$ we have that for photons one has to set $k=0$. From the Hamilton-Jacobi formalism we obtain the action $S$ in the form 
\begin{eqnarray}\label{separation}
S= -Et+L\varphi+S_{r}(r)\, ,
\end{eqnarray}
where $E$ and $L$ are the usual conserved quantities for the energy and angular momentum of the photon, respectively and  $S_{r}$ is function of $r$ only. Now it is straightforward to obtain the Hamilton-Jacobi equation in the following form
\begin{eqnarray}
\label{Eq:Sep1}
k&=&-\frac{E^2}{F(r)}+F(r)
\left(\frac{\partial S_{r}}{\partial r}\right)^2
+ \frac{L^2}{r^2} \, .\nonumber\\
\end{eqnarray}
Then from the separability of the action given in  Eq.~(\ref{Eq:Sep1}) we obtain the radial component of equations of motion for photons in the following form
\bea \label{Veff3}
\dot{r}^2=E^2-\frac{L^2}{r^2}F(r)
\, .
\eea
To find the radii of circular orbits for given values of $E$ and $L$ we can then solve simultaneously $\dot{r}=\ddot{r}=0$, i.e., 
\begin{eqnarray}
\tilde{V}_{eff}(r,E,L)=0, \mbox{~~~} \frac{\partial \tilde{V}_{eff}(r,E,L)}{\partial r}=0\, ,
\end{eqnarray}
where the function $\tilde{V}_{eff}(r,E,L)$ is defined as
\begin{eqnarray}
\tilde{V}_{eff}(r,E,L)&=& E^2-\frac{L^2}{r^2}F(r)\, .
\end{eqnarray}
In (3+1)-dimensional spacetime, for photon sphere one needs to solve $\tilde{V}_{eff}'=0$, but interestingly it is not sufficient to find the photon sphere around (2+1)-dimensional BTZ black hole in the GB gravity. Therefore one may determine it using additional condition $\tilde{V}_{eff}''=0$. Here, this condition gives $r_{ph}$ implicitly as
\begin{eqnarray} 
3 (4 \alpha  \Lambda -1)r^2 +8 \alpha  m =0\, .
\end{eqnarray} 
 
\begin{figure*}[t!]
\begin{center}
\includegraphics[width=0.49\linewidth]{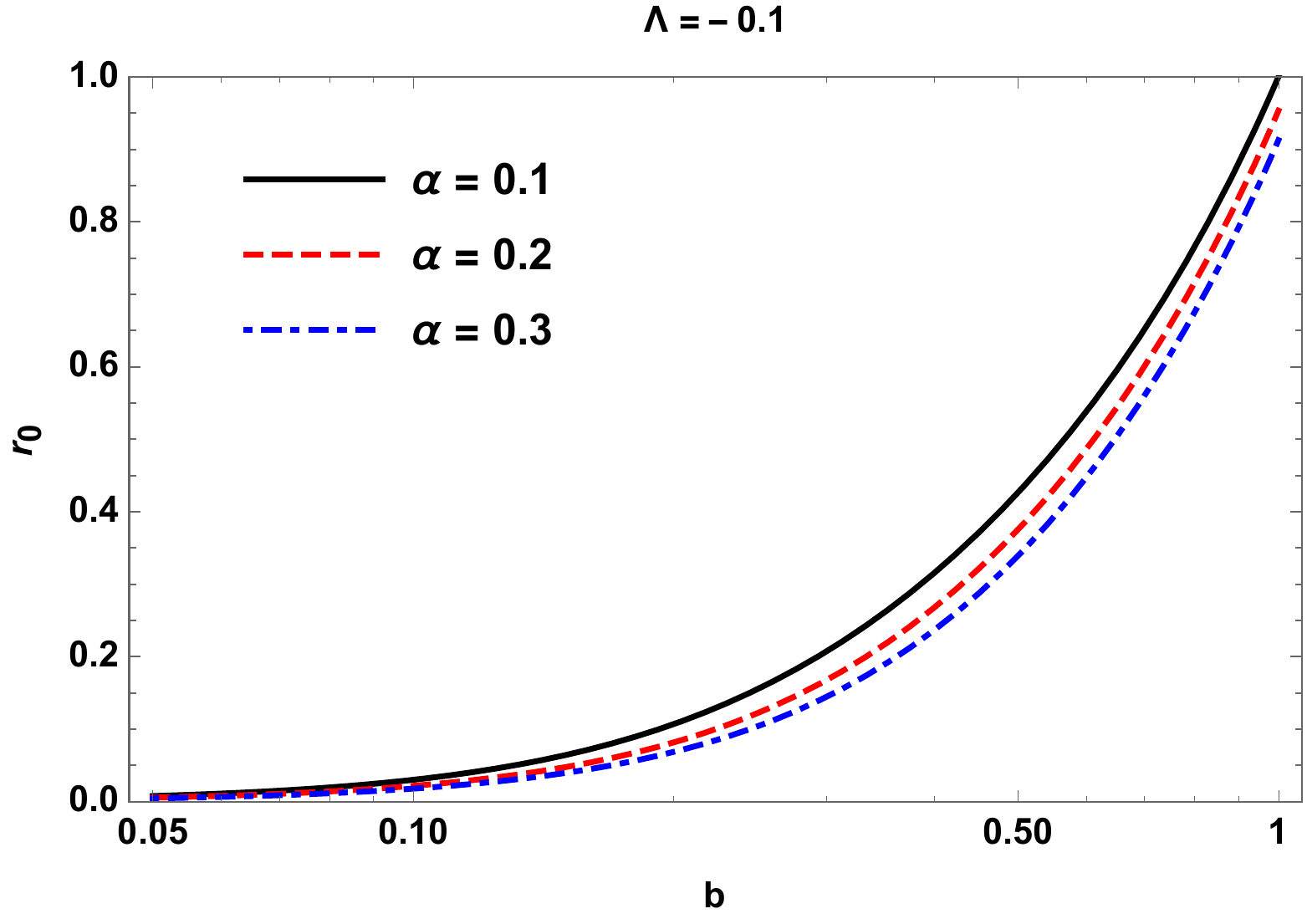}
\includegraphics[width=0.49\linewidth]{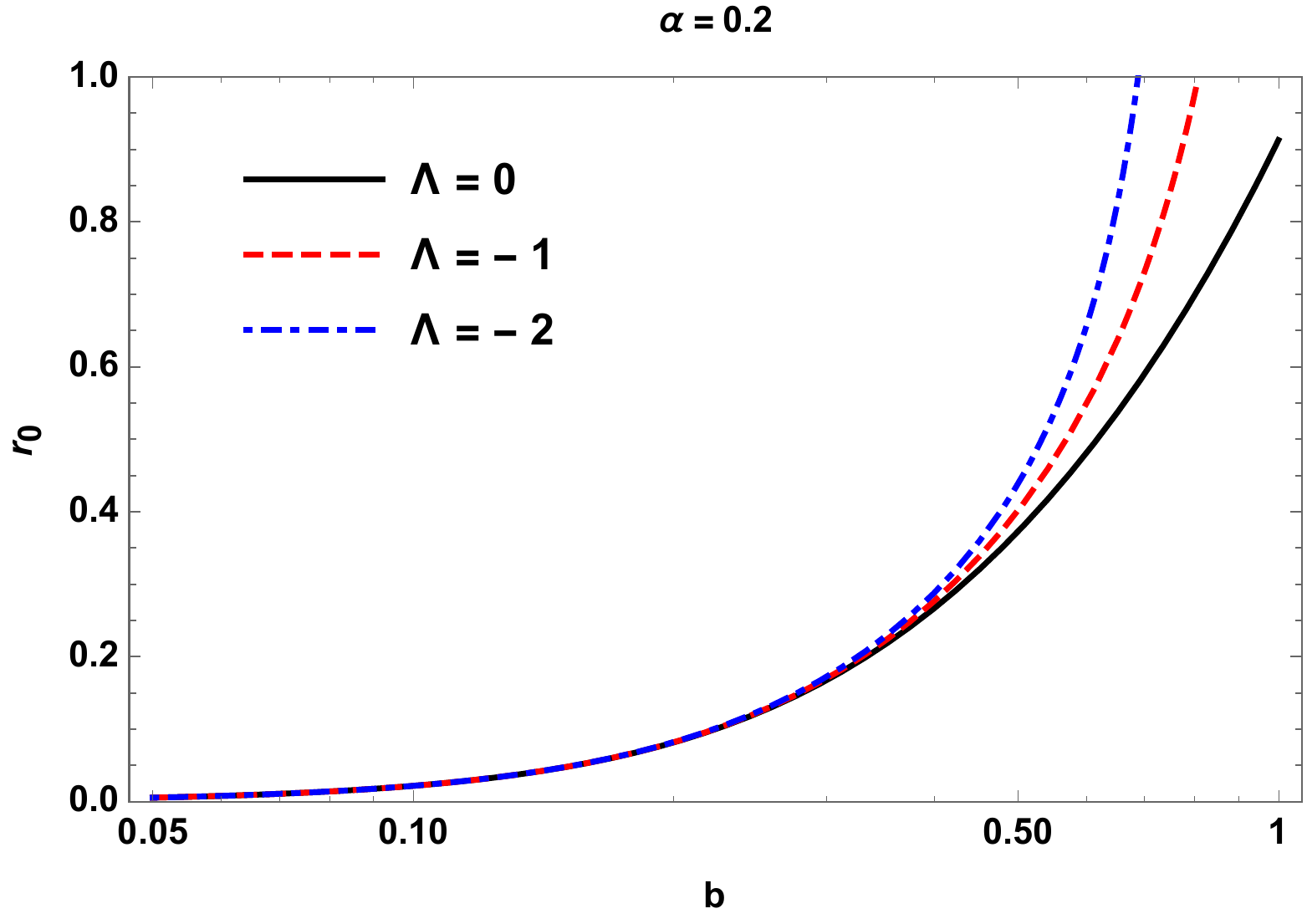}
\end{center}
\caption{The dependence of the distance of closest approach of photon on the impact parameter $b$. \label{rr}}
\end{figure*}

In the limit of $\alpha \Lambda \ll 1$ we can write the approximate expressions for the photon orbit $r_{ph}$ as
\begin{eqnarray}\label{Eq:rph}
r_{ph}&=& \frac{2 \sqrt{6}}{3}  \left(1+2 \alpha  \Lambda \right)\sqrt{\alpha m } +O(\Lambda^2) \, .
\end{eqnarray}
This clearly shows the parameter $\alpha$ increases the radius of the photon orbits. 
From the equation of motion for the mass-less particle one can write $u_\mu=\frac{\partial S}{\partial x^\mu}$ (and $u^\mu=g^{\mu\nu}u_\nu$) that allows to write explicit form of the components of the "three"-velocity as
\begin{eqnarray}
\dot{t}&=&\frac{2 \alpha}{r^2 \left(\sqrt{-4 \alpha \Lambda-\frac{4 \alpha m}{r^2}+1}-1\right)} E\ ,\label{t}\\
\dot{\phi}&=&\frac{L}{r^2}\ ,\label{phi}\\
\dot{r}^2&=&E^2+\frac{L^2 \left(1-\sqrt{1-4 \alpha \Lambda-\frac{4 \alpha m}{r^2}}\right)}{2 \alpha}\ ,\label{r}
\end{eqnarray}
where differentiation is made over the affine parameter. Let us introduce so called the distance of closest approach $r_0$ defined as the minimum distance between the central gravitating object and the trajectory of photons. From the equation (\ref{r}) one can obtain the value for this distance that reads

\begin{eqnarray}
r_0=\frac{L^2 \sqrt{m}}{\sqrt{\alpha E^4+\Lambda L^4+E^2 L^2}}\ .
\end{eqnarray}
The dependence of the distance of closest approach on the impact parameter $b=L/E$ is plotted in Fig.~\ref{rr}. It is apparent from the figure that the increase of the GB parameter decreases the distance of closest approach slightly while the increase of the absolute value of the cosmological constant shows opposite behaviour. It can be interpreted as follows, since the increase of the negative value of cosmological constant increases the repulsive force between the central object and any particle with positive energy then the increase of such repulsive force should in turn increase the closest distance between the trajectory of a particle and the central object for the given impact parameter $b$ which is shown in the right panel of Fig. \ref{rr}.

Let us now investigate the bending angle of a photon moving in the spacetime of the BTZ black hole in the GB gravity. To do so one can introduce new variable as $u=1/r$ that simplifies our calculation to find the bending angle. One can write the equation of motion in terms of new variable as
\begin{eqnarray}
\frac{d\phi}{du}=\frac{L}{\sqrt{E^2+\frac{L^2 \left(1-\sqrt{1-4 \alpha \Lambda-4 \alpha m u^2}\right)}{2 \alpha}}}
\end{eqnarray}

The resultant bending angle then takes the following form

\begin{eqnarray}
\delta=2 \int_{0}^{u_0}\frac{L}{\sqrt{E^2+\frac{L^2 \left(1-\sqrt{1-4 \alpha \Lambda-4 \alpha m u^2}\right)}{2 \alpha}}} du-\pi\ .
\end{eqnarray}

One can explore the case  when the GB parameter $\alpha$ is small (i.e. $\alpha \Lambda \ll1$) and use the linear approximation in it. In this approximation the integral above can be written as
\begin{widetext}
\begin{eqnarray}
\delta=2 \int_{0}^{u_0}\frac{L}{\sqrt{\Lambda L^2+L^2 m u^2+E^2}}-\frac{\alpha L^3 \left(\Lambda+m u^2\right)^2}{2 \left(\Lambda
	L^2+L^2 m u^2+E^2\right)^{3/2}} du-\pi\ .
\end{eqnarray}

After integration the bending angle becomes

	\begin{eqnarray}
	\delta&=&\frac{\frac{\left(L^2 (4-\alpha \Lambda)+3 \alpha
			E^2\right) \log \left(\sqrt{m} \sqrt{L^2 \left(\Lambda+m
				u_0^2\right)+E^2}+L m u_0\right)}{\sqrt{m}}-\frac{\alpha L u_0
			\left(E^2 L^2 \left(2 \Lambda+m u_0^2\right)+\Lambda L^4 \left(\Lambda+m
			u_0^2\right)+3 E^4\right)}{\left(\Lambda L^2+E^2\right) \sqrt{L^2
				\left(\Lambda+m u_0^2\right)+E^2}}}{2 L^2}
	\\\nonumber
	&-&\frac{\left(L^2 (4-\alpha \Lambda)+3 \alpha E^2\right) \log \left(\sqrt{m} \sqrt{\Lambda
			L^2+E^2}\right)}{2 L^2 \sqrt{m}}-\pi\ .
	\end{eqnarray}
\end{widetext}

\begin{figure*}[t!]
	\begin{center}
		\includegraphics[width=0.49\linewidth]{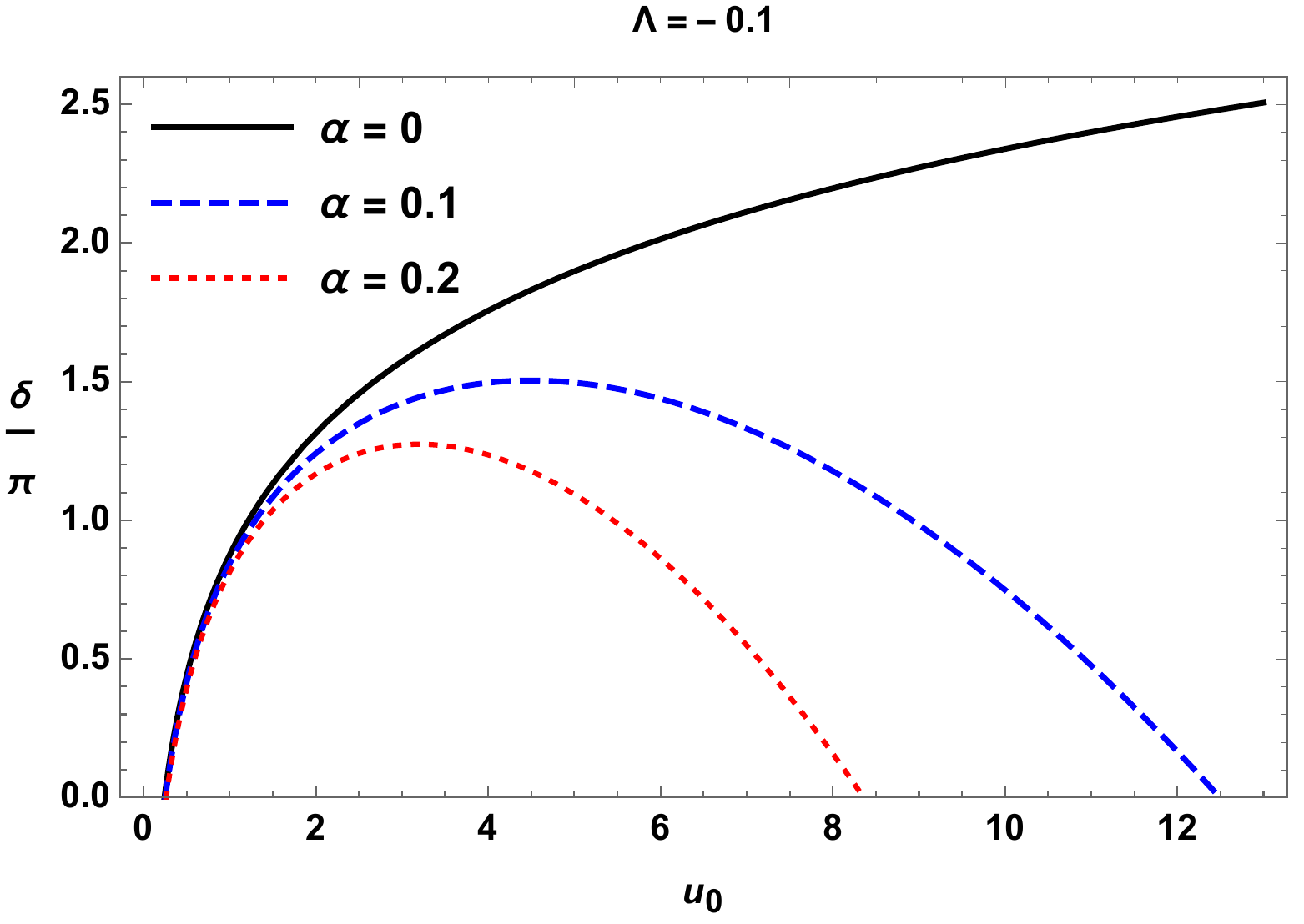}
		\includegraphics[width=0.49\linewidth]{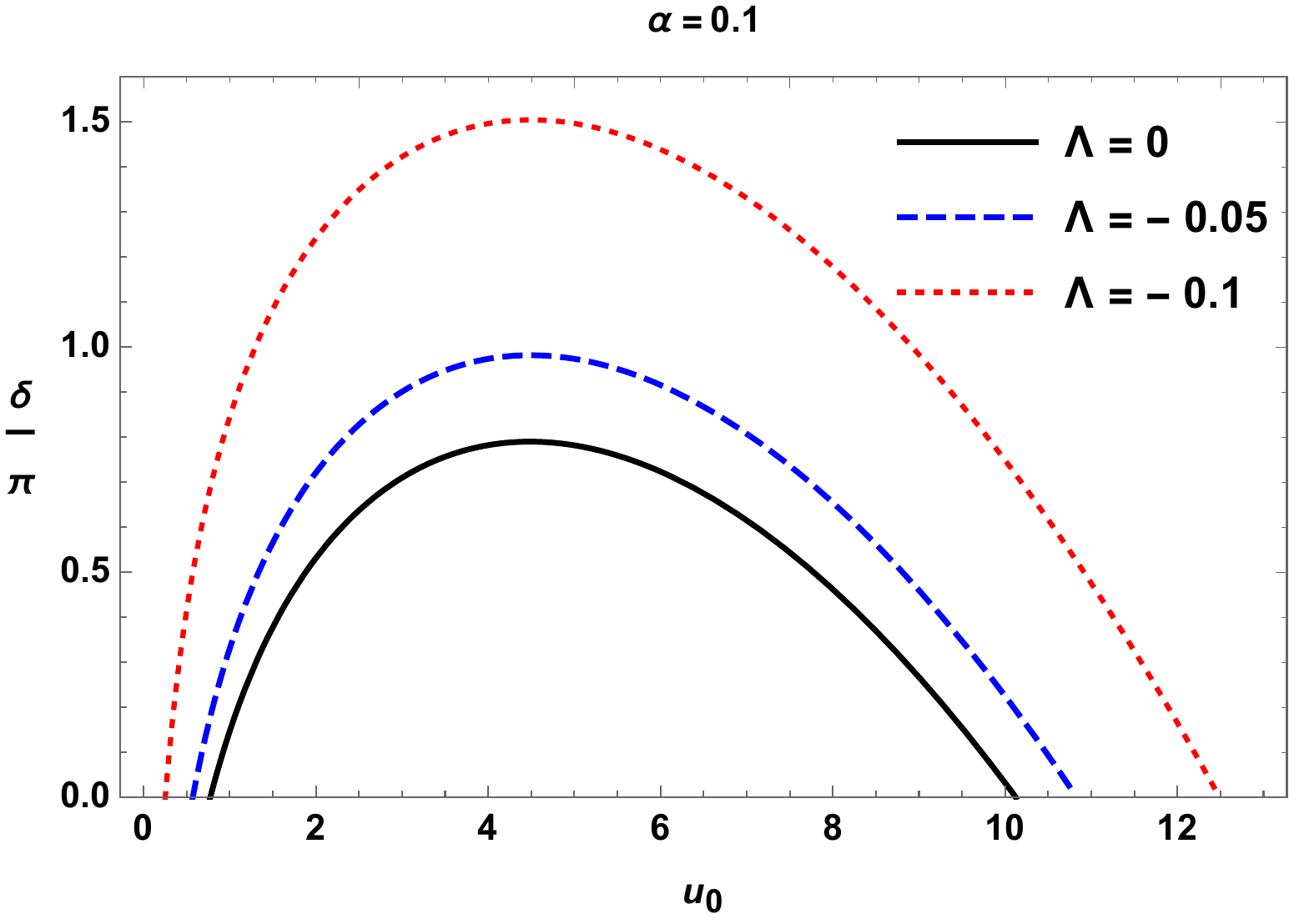}
	\end{center}
	\caption{The dependence of the bending angle of photon on the inverse of the distance of closest approach $u_0$. \label{d}}
\end{figure*}

The dependence of such bending angle on the parameter $u_0$ is plotted in Fig.~\ref{d} for the chosen energy $E=1$ and angular momentum $L=3$ of the photon and different values of the GB parameter $\alpha$ with cosmological constant $\Lambda$.

One can easily see from the left figure that in the absence of GB parameter the bending angle increases monotonically when one increases the inverse of closest approach distance, $u_0$. When this parameter comes to the game we see that the bending angle increases until its peak first and then starts to decrease and becomes zero for some value of $u_0$. In this case the maximum deflection angle can be easily found by setting equal to zero the derivative of the deflection angle over $u_0$ which in turn reduces to the equation

\begin{eqnarray}
\frac{2 E^2 L-L^3 \left(\Lambda+m u_0^2\right) \left(\alpha \left(\Lambda+m
	u_0^2\right)-2\right)}{\left(L^2 \left(\Lambda+m
	u_0^2\right)+E^2\right)^{3/2}}=0\ .
\end{eqnarray}

The maximum point then becomes

\begin{eqnarray}
u_0^*=\left[\frac{\sqrt{2 \alpha E^2+L^2}}{\alpha L m}+\frac{1}{\alpha m}-\frac{\Lambda}{m}\right]^\frac{1}{2}\ .
\end{eqnarray}

\section{Conclusions}
\label{Sec:conclusion}
In the present study we have seen that the GB parameter has opposite effect on the radii of stable circular orbits with respect to the cosmological constant. The GB parameter allows particles not to be on stable circular orbits while the cosmological constant restores the stable orbits for particles around the BTZ black hole in the 3-D EGB gravity. It is interesting that in the case of $\Lambda=0$  there occurs no stable circular orbits around the BTZ black hole in the 3-D EGB gravity. In other words the particle under the effect of $\alpha$ alone either can escapes to infinity or falls into the black hole. The situations however gets overturned once the effect arising from the cosmological constant is taken into consideration. It thus appears that cosmological constant in (2+1)-dimensional BTZ black hole spacetime in the GB gravity plays a crucial role for the existence of stable circular orbits for particles around the black hole. Also note that the ISCO radius increases with increasing the GB parameter while it gets decreased with increasing the cosmological constant.

Investigation of photon motion around the BTZ black hole in the 3-D GB gravity shows that the  radius of the photon orbit increases with the increase in the GB parameter while the cosmological constant with its negative value decreases it. Study of the bending angle of the photon approaching the central object from infinity shows that in the presence of the cosmological constant its value increases monotonically in the absence of the GB parameter while in the presence of latter it reaches its peak corresponding to the specific value of the parameter $u_0^*$ and then goes down. It also has been shown that the increase of the GB parameter reduces the bending angle while the increase in the absolute value of the negative cosmological constant makes such angle bigger.

\textcolor{black}{In a recent work Hennigar et. al \cite{Hennigar20a}, have obtained the charged and rotating black hole solutions in the novel 3D
GB theory of gravity which is generalization of the BTZ solution. Their charged metric is obtained in the Maxwell and Born–Infeld theories. In a separate work we will analyse the motion of charged particles and photons in these newly obtained charged and rotating spacetimes in the 3D GB gravity, to look at the effects of the charge and rotation on the motion of particles.}

\section*{Acknowledgments}
B.N. acknowledges support from the China Scholarship Council (CSC), grant No. 2018DFH009013. This research
is supported by Grants of the Uzbekistan Ministry for Innovative Development and by the Abdus
Salam International Centre for Theoretical Physics under Grant No. OEA-NT-01.

\appendix

\bibliographystyle{apsrev4-1}  
\bibliography{gravreferences}

\end{document}